\documentclass[twocolumn,showpacs,preprintnumbers,amsmath,amssymb,superscriptaddress]{revtex4}

\usepackage{graphicx}
\usepackage{dcolumn}
\usepackage{bm}

\begin{document}

\preprint{APS/123-QED}

\title{A Neutron Scattering Study of the \textit{H-T} Phase Diagram of the Bond Frustrated Magnet ZnCr$_2$S$_4$}

\author{D.~Hsieh}
\affiliation{Joseph Henry Laboratories of Physics, Princeton University, Princeton, New Jersey 08544, USA}
\affiliation{(Present Address) Institute for Quantum Information and Matter, California Institute of Technology, Pasadena, California 91125, USA}
\author{Y.~W.~Li}
\affiliation{Joseph Henry Laboratories of Physics, Princeton University, Princeton, New Jersey 08544, USA}
\author{S.~Watauchi}
\affiliation{Department of Chemistry, Princeton University, Princeton, New Jersey 08544, USA}
\author{Chang~Liu}
\affiliation{Joseph Henry Laboratories of Physics, Princeton University, Princeton, New Jersey 08544, USA}
\author{Q.~Huang}
\affiliation{NIST Center for Neutron Research, Gaithersburg, Maryland 20899, USA}
\author{J.~W.~Lynn}
\affiliation{NIST Center for Neutron Research, Gaithersburg, Maryland 20899, USA}
\author{R.~J.~Cava}
\affiliation{Department of Chemistry, Princeton University, Princeton, New Jersey 08544, USA}
\author{M.~Z.~Hasan}
\affiliation{Joseph Henry Laboratories of Physics, Princeton University, Princeton, New Jersey 08544, USA}

\date{\today}

\begin{abstract}
Detailed neutron scattering measurements on the bond frustrated
magnet ZnCr$_2$S$_4$ reveal a rich $H$-$T$ phase diagram. The field
dependence of the two subsequent antiferromagnetic transitions
follows closely that of recently reported structural instabilities,
providing further evidence for spin driven Jahn-Teller physics. The
incommensurate helical ordered phase below $T_{N1}=$ 15.5 K exhibits gapless
spin wave excitations, whereas a spin wave gap $\Delta\simeq$ 2 meV
opens below $T_{N2}=$ 8 K as the system undergoes a first-order transition to a commensurate collinear
ordered phase. The spin wave gap is closed by a strong magnetic
field.
\end{abstract}

\maketitle

Strong geometrical frustration is consistently being found in
magnetic $B$-site spinel oxides $AB_2X_4$ ($X$=O)
\cite{Ramirez,Lee,Lee-V,Chung,Ueda,Matsuda}. These systems are
characterized by Heisenberg spins on a lattice of corner sharing
tetrahedra interacting via antiferromagnetic (AFM) nearest-neighbor
direct exchange. In the spinel chalcogenides ($X$=S, Se) on the
other hand, the nearest-neighbor direct exchange is weakened due to
a larger $B$-$B$ distance and the small ferromagnetic (FM) Cr-S-Cr
superexchange becomes the dominant interaction between them.
Nevertheless the spins often remain frustrated because competing
further neighbor exchanges become important
\cite{Hemberger,Rudolf,Tsurkan,Menyuk}. Strong spin-phonon coupling
plays a major role in relieving both geometrical and bond
frustration in these systems, manifested by both uniform and
nonuniform lattice distortions that commonly accompany the magnetic
ordering \cite{Hidaka-zerofield,Hidaka-field}. Characterizing these distortions in bond frustrated
spinels is particularly crucial in light of their recently
discovered novel magnetoelectric properties
\cite{Cheong,Hemberger-nature,Gnezdilov}. While the structural
distortions involved are small and largely nonuniform, thus making
them difficult to detect using standard diffraction techniques, its
effect on system dynamic variables such as phonons and magnons can
be clearly measured.

The correlated insulator ZnCr$_2$S$_4$ is a strongly bond frustrated material
(the value of the frustration parameter $f$ = $|\Theta_{CW}|/T_N
\simeq$ 0.5). Hamedoun \textit{et al.} performed neutron diffraction
measurements in zero magnetic field and reported the following
results \cite{Hamedoun-neutron}. At $T_{N1}$=15.5 K, the material
undergoes a second order phase transition from a paramagnetic to a
helical ordered phase. The helical structure is characterized by an
incommensurate ordering wave vector $\vec{k}_1$=(0 0 0.79) with
spins confined and ferromagnetically ordered within the (001)
planes. Below $T_{N2}$=8 K, the helical structure partially
transforms into two collinear commensurate structures with
$\vec{k}_2$=($\frac{1}{2}\frac{1}{2}$0) and
$\vec{k}_3$=(1$\frac{1}{2}$0) with spins oriented parallel to the
[1$\bar{1}$0] and [001] directions respectively. More recently,
specific heat, thermal expansion and phonon anomalies have been
reported at both $T_{N1}$ and $T_{N2}$, and an external magnetic
field has been found to suppress only the anomalies at $T_{N2}$
\cite{Hemberger}. These results suggest that the helical structure
is stabilized over the collinear structures in a field, and that
spin-phonon coupling may be driving structural distortions at both
$T_{N1}$ and $T_{N2}$ that have so far not been resolved. In this
Letter, we use neutron scattering to study the magnetic field
dependence of both the spin order and dynamics of ZnCr$_2$S$_4$ for
the first time. Our results show that the magnetic order parameters
display the same field dependence as the reported structural
anomalies do, providing stronger support for spin-driven structural
instabilities at both $T_{N1}$ and $T_{N2}$.

A 10 g powder sample of ZnCr$_2$S$_4$ was prepared by solid state
reaction between stoichiometric amounts of ZnS, Cr and S. Rietveld
refinement of neutron diffraction data from the NIST Center for
Neutron Research BT-1 diffractometer showed that the sample is
single phase spinel with space group $Fd\bar{3}m$ with lattice
parameter $a$=9.986 \AA\ and sulphur fractional coordinate $u$=
0.385 for $T$=300 K. Magnetic properties were measured using a
commercial SQUID magnetometer (Quantum Design MPMS-5). Neutron
diffraction scans were performed at NIST on the thermal-neutron
triple-axis spectrometer BT-7 using a fixed incident and final
energy of 14.7 meV \cite{BT7}. We used PG(002) reflections for both the
monochromator and the analyzer which was operated in flat mode, and
used horizontal beam collimations of open-50'-50'-open. Elastic and
inelastic neutron scattering measurements were performed on the cold
neutron time-of-flight spectrometer (DCS) using 3 \AA\ and 4.8 \AA\
incident neutrons. All magnetic fields were applied perpendicular to
the scattering plane. We used the integrated intensity of several
nuclear Bragg peaks to determine the normalized magnetic neutron
scattering intensity \cite{Lovesey}

\begin{align}
\tilde{I}(Q,\omega) =
&\int\frac{d\Omega_{\hat{Q}}}{4\pi}\left|\frac{g}{2}F(Q)\right|^2\\
\notag
&\times\sum_{\alpha\beta}(\delta_{\alpha\beta}-\hat{Q}_{\alpha}\hat{Q}_{\beta})S^{\alpha\beta}(\vec{Q},\omega)
\end{align} where $F(Q)$ is the magnetic form factor for
Cr$^{3+}$, $S^{\alpha\beta}(\vec{Q},\omega)$ is the dynamic spin
correlation function and $g$ is the Land\'{e} $g$-factor. Error bars where indicated are statistical in origin and represent one standard deviation.

\begin{figure}
\includegraphics[width=7cm]{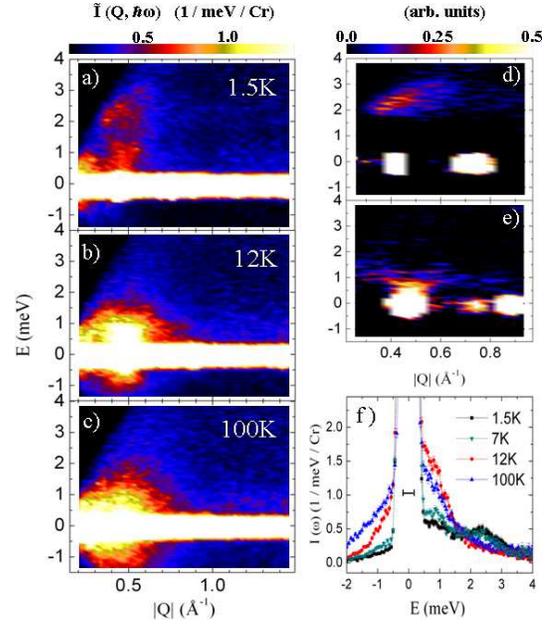}
\caption{\label{fig:DCS_T} \textbf{Zero field magnetic excitations:}
Inelastic neutron scattering spectra at zero field and (a) $T$=1.5
K, (b) $T$=12 K and (c) $T$=100 K, taken using 3 \AA\ incident
neutrons. The dispersion of the magnetic excitations are more
clearly seen in the difference spectra (d) $\tilde{I}$(1.5
K)$-\tilde{I}$(12 K) and (e) $\tilde{I}$(12 K)$-\tilde{I}$(100 K).
(f) $\tilde{I}$($Q$,$\omega$) averaged over 0.3 \AA$^{-1}$$<Q<$0.6
\AA$^{-1}$. Horizontal bar shows instrument resolution.}
\end{figure}

Figures~\ref{fig:DCS_T}(a)-~\ref{fig:DCS_T}(c) are color images of
$\tilde{I}(Q,\omega)$ at three temperatures. In the paramagnetic
phase, Fig.\ref{fig:DCS_T}(c) shows evidence for critical
fluctuations of small helically correlated clusters. Here, the
static spin correlation length $\xi$ is close to the Cr-Cr
separation as determined by neutron diffraction. Strong fluctuations
of small AFM clusters at $T\gg T_N$ is also a hallmark of
geometrically frustrated magnets. At $T$=12 K, in the long-range
helical ordered phase, rather well defined spin wave excitations
appear (Figs.1(b) and (e)) whose energy tends to zero at the
magnetic Brillouin zone center $k_1$, and reaches a zone boundary
energy near 1.5 meV at $Q\sim$0.3 \AA$^{-1}$. By subtracting the
nuclear incoherent scattering from $\tilde{I}(Q,\omega)$ and
integrating over $\hbar\omega$ and $Q$, we obtained the sum rule
$S(S+1)$=3.8(3) at 12 K, which is the expected value of 15/4 = 3.75 for
orbitally-quenched Cr$^{3+}$ ($S$=3/2) ions. This and the $Q$
dependence tell us that the scattering is magnetic. By integrating
the 12 K data over $\hbar\omega\in[0.5,3]$meV and
$Q\in[0.2,1]$\AA$^{-1}$ we obtain the total fluctuating moment
$\langle \delta m \rangle^{2}=(3/2)\int \hbar d\omega\int Q^{2}dQ
[\tilde{I}(Q,\hbar\omega)/|F(Q)|^{2}]/\int Q^{2}dQ$ =1.20(3)/Cr,
which is considerably less than that found in ZnCr$_2$O$_4$ at the
lowest temperatures \cite{Lee-PRL}. For $T<T_{N2}$
(Fig.~\ref{fig:DCS_T}(a)), $\langle \delta m \rangle^{2}$ decreases
slightly to 0.96(7)/Cr, but nearly half of this spectral weight
shifts to higher energies between 2$-$3 meV.
Figure~\ref{fig:DCS_T}(f) shows the energy dependence of the
magnetic scattering at $Q$=0.45 \AA$^{-1}$ averaged over an 0.3
\AA$^{-1}$ interval. Even below $T_{N2}$, intensity continues to be
transferred from a low energy maximum centered about 1 meV to a high
energy maximum centered around 2.5 meV. This and the difference of
the $T$=1.5 K and $T$=12 K spectra (Fig.~\ref{fig:DCS_T}(d))
strongly suggest that the low energy scattering is due to remnant
helically ordered regions of the sample. The excitations intrinsic to
the collinear phase are seen to be gapped by approximately 2 meV and
disperse with a bandwidth ($\sim$1.5 meV) close to that observed in
the helical phase. The latter fact indicates that any structural
distortion occuring at $T_{N2}$ does not alter the exchange
constants significantly. However the origin of the gap is unclear,
one possibility being some single-ion anisotropy that develops via a
local distortion of the Cr-S octahedra. Alternatively, more exotic
scenarios such as a lifted local resonance mode \cite{Lee-PRL} may
be at play.

\begin{figure}
\includegraphics[width=8cm]{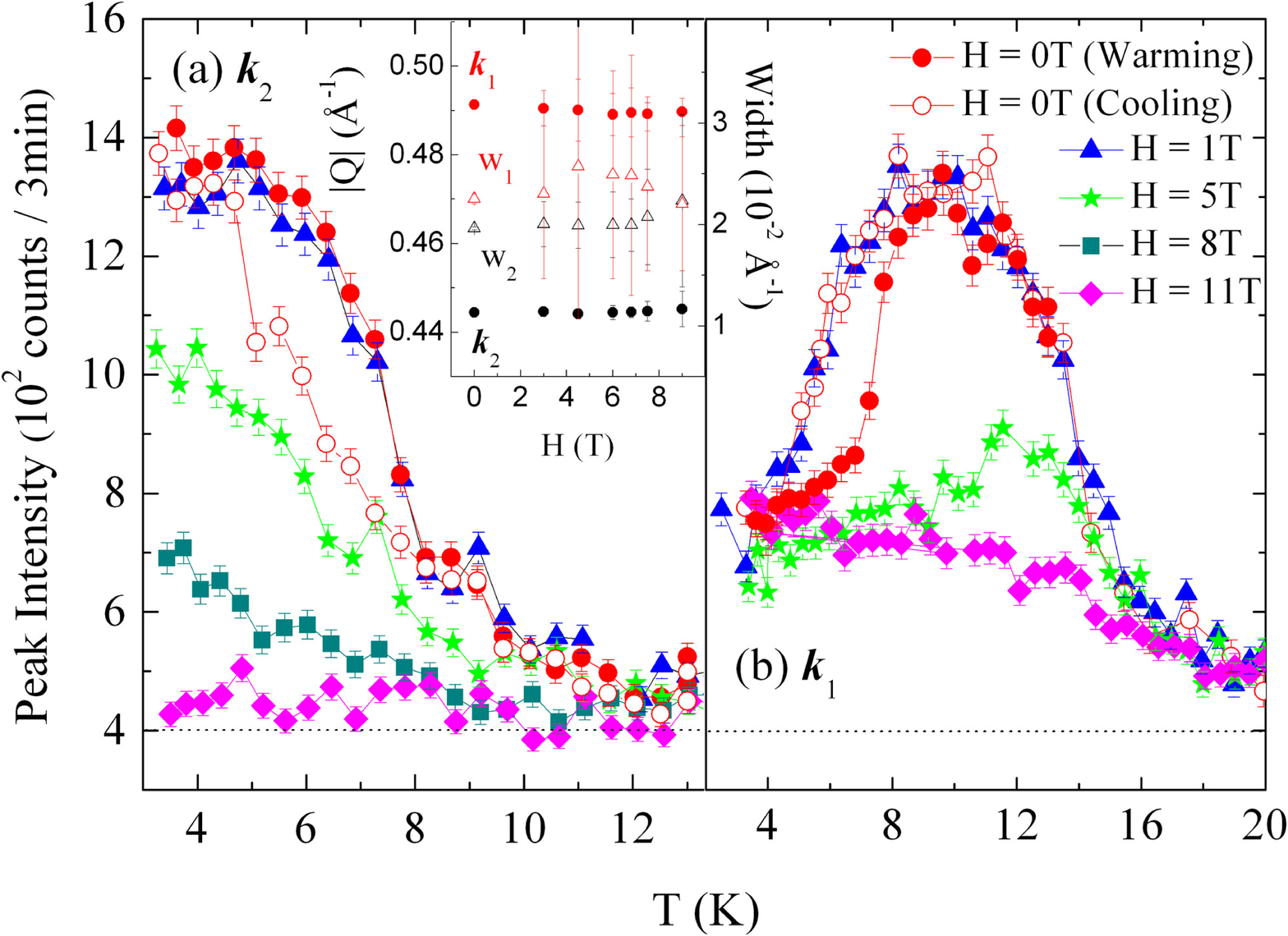}
\caption{\label{fig:Elastic_T} \textbf{Order parameter behavior:}
Temperature dependence of the (a) $\vec{k}_2$ and (b) $\vec{k}_1$
peak intensities in various external fields taken on BT-7 upon
cooling and warming respectively. Both warming and cooling curves
are displayed for the $H$=0 T data to show thermal hysteresis. The
dotted line indicates the background (BG) level due to nuclear
incoherent scattering. The inset of (a) shows the field dependence
of the $\vec{k}_1$ and $\vec{k}_2$ peak positions and widths at
$T$=4 K determined by fitting Gaussians to DCS data.}
\end{figure}

We proceed to study the magnetic field dependence of the phase
boundary between the ordered magnetic states. The inset of
figure~\ref{fig:Elastic_T}(a) shows that both the $\vec{k}_1$ and
$\vec{k}_2$ Bragg peak positions stay fairly constant and remain
resolution limited ($\xi>$ 50 \AA) from $H$=0 T up to $H$=9 T,
therefore the peak intensity $I$ is a good measure of the order
parameter. Figure~\ref{fig:Elastic_T} shows the temperature
dependence of both peak intensities under various external fields. In
zero field, $I(\vec{k}_1)$ develops long range order near $T$=15 K, has a rounded
maximum at $T$=10 K, and then falls off sharply to a nearly constant
non-zero value below $T$=6 K. The downturn in $I(\vec{k}_1)$
coincides with a development of long range order associated with $I(\vec{k}_2)$, and signals the
partial transformation from helical to collinear order. Both peaks
exhibit clear thermal irreversibility around $T_{N2}$ which is
consistent with bulk data \cite{Hemberger}. The intensity difference
between the warming and cooling curves at $T$=7 K remained unchanged
out to a waiting time of 60 minutes, which strongly suggests that
the thermal irreversibility results from true hysteresis associated
with a first order transition. Previous work has shown that the
ratio $R=I(\vec{k}_1)$/$I(\vec{k}_2)$ at the lowest temperature is
positively correlated with the number of sulphur vacancies in the
sample \cite{Hamedoun-neutron}. Our measured value of $R$=0.35(3) at
$T$=4 K corresponds well to previous results on sulphur annealed
samples.

\begin{figure}
\includegraphics[width=8cm]{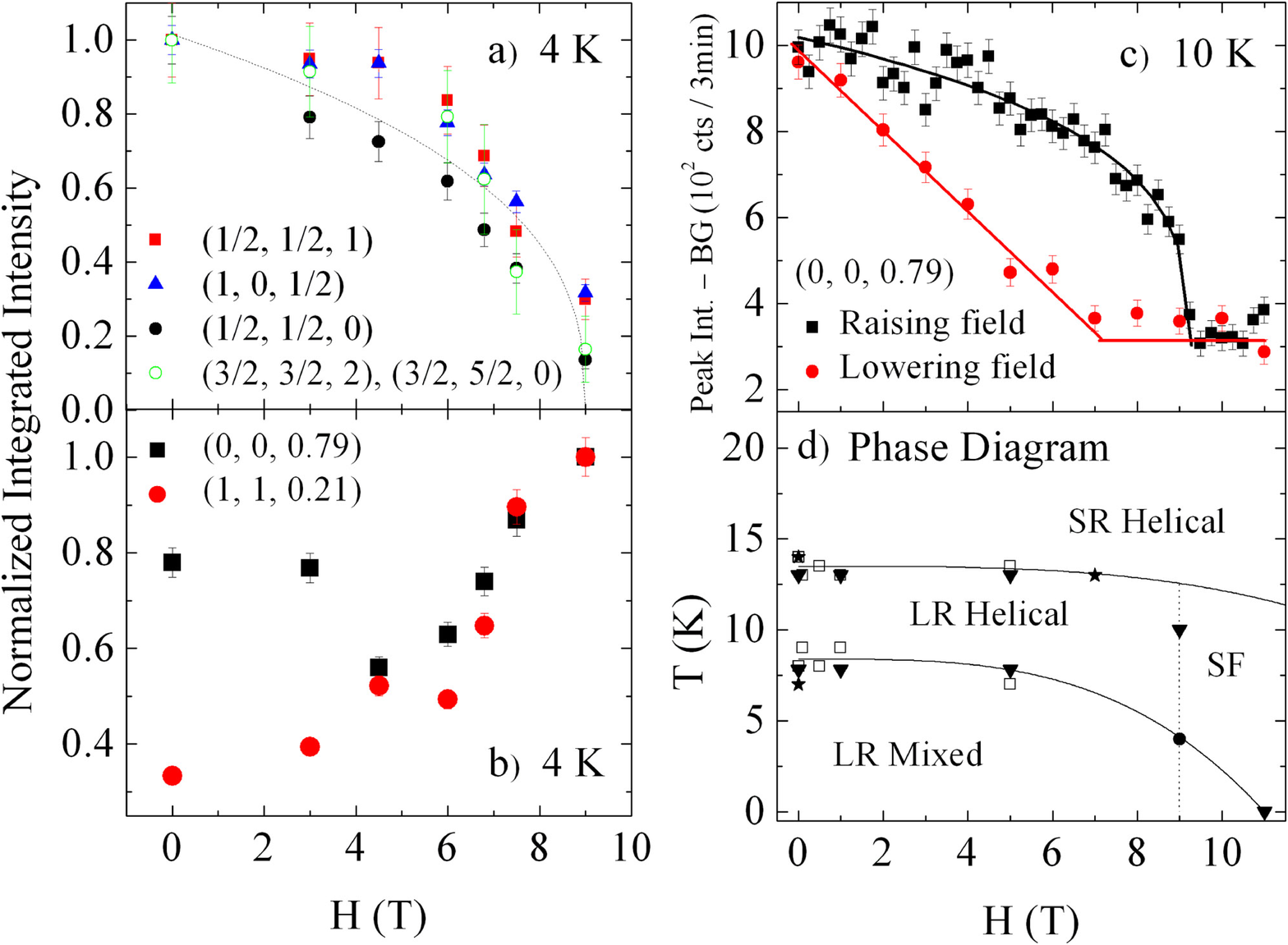}
\caption{\label{fig:Phase_Diagram} \textbf{Determination of phase
diagram:} Magnetic field dependence of the (a) $\vec{k}_3$,
$\vec{k}_2$, (001)$^{\vec{k}_2}$, (112)$^{\vec{k}_2}$ and
(120)$^{\vec{k}_2}$ and (b) $\vec{k}_1$ and (111)$^{-\vec{k}_1}$
integrated intensities at 4K normalized by their zero field values
and $H$=9 T values respectively (data from DCS). (c) Background
subtracted intensity of the $\vec{k}_1$ peak taken on BT-7 at $T$=10
K upon raising followed by lowering the magnetic field. (d) $H-T$
phase diagram of ZnCr$_2$S$_4$ constructed using the following data
sets: ($\square$) Peak in $d(\chi T)/dT$, ($\star$) peak in
$C_p(T)$, ($\blacktriangledown$) 3 axis neutrons, ($\bullet$) TOF
neutrons. Dashed line is the approximate spin-flop (SF) boundary.
All lines are guides to the eye.}
\end{figure}

As the field is raised, we observe a decrease in the onset
temperature of the $\vec{k}_2$ phase and an overall decrease in
$I(\vec{k}_2)$ (Fig.~\ref{fig:Elastic_T}(a)). At $T$=4 K,
figure~\ref{fig:Phase_Diagram}(a) shows that higher order peaks
belonging to the complex collinear phase all decay with a constant
relative intensity before disappearing at a critical field
$H_c$(4K)$\sim$9 T. A suppression of both thermal and phonon
anomalies at $T_{N2}$ has also been reported in fields of 7 T
\cite{Hemberger}. These results show that the long-range ordering
temperature of the collinear phase of ZnCr$_2$S$_4$ is suppressed to
zero by a field close to its mean field energy scale
$k_BT_{N2}/g\mu_B\sim$ 6 T.

In contrast, the onset temperature of long-range helical order does
not change significantly with field (Fig.~\ref{fig:Elastic_T}(b)).
This is consistent with specific heat measurements which show little
change in both the position and magnitude of the anomaly at $T_{N1}$
up to $H$=7 T \cite{Hemberger}. Therefore unlike the collinear
structure, the helical structure is not being suppressed by the
applied field. The shape of the $I(\vec{k}_1)$ versus $T$ curve does
however undergo a qualitative change. At $H$=5 T, $I(\vec{k}_1)$
rises below $T_{N1}$ to a sharp maximum near $T$=13 K with a
diminished amplitude relative to the zero field case, and then
decreases gradually upon further cooling. By $H$=11 T, this maximum
completely disappears and we instead observe a monotonic increase of
intensity with cooling. The intensity at the lowest temperature,
which arises from residual helical regions that are pinned by
impurities, is fairly independent of field. We now consider several
possibilities to explain the suppression of $I(\vec{k}_1)$. (i) If spin
anisotropy is present, external fields can exert a torque on
magnetically ordered powder grains causing $\vec{k}_1$ to rotate
away from the $\hat{Q}$ direction. This can be ruled out because no
changes were observed in the (220), (311) and (004) nuclear Bragg
reflections with field. Moreover, the zero field values of
$I(\vec{k}_1)$ were recovered upon removing the field. (ii) If the
magnetic field is large enough to overcome all internal AFM exchange
energies, all spins can be made to align with the field. This can
again be ruled out because no ferromagnetic contribution was
detected atop the nuclear Bragg peaks, and because the bulk
magnetization at $H$=11 T is far below saturation \cite{Ueda}. (iii)
The possibility of new magnetic structures induced by the external
field is ruled out based on the absence of new diffraction peaks.
(iv) The magnetic field may induce a spin-flop transition to an
arrangement where spins preserve (0 0 0.79) ordering but orient
perpendicular to $\vec{H}$ to take advantage of canting. Since
$\tilde{I}(Q,0)$ is proportional to the component of the ordered
moment perpendicular to $\hat{Q}$, this would naturally explain the
decrease in $I(\vec{k}_1)$.

The spin-flop field $H_{SF}$ in the purely helical phase lies
between 7 T and 9 T as shown in figure~\ref{fig:Phase_Diagram}(c),
which is slightly higher than previous estimates ($H_{SF}\sim$5 T)
based on single crystal magnetization data
\cite{Hamedoun-magnetization}. Upon entering the mixed phase,
$I(\vec{k}_1)$ becomes rather insensitive to field as shown in
figure~\ref{fig:Elastic_T}(b). However a general positive
correlation is evidenced in figure~\ref{fig:Phase_Diagram}(c), which
suggests that an external field stabilizes the helical structure
over the collinear structure. The fact that the relative intensity
of the $\vec{k}_1$ and (111)$^{-\vec{k}_1}$ reflections does not
stay constant is further support for a spin-flop transition. The
magnetic phase boundaries deduced from our neutron scattering
experiments, along with complementary dc-susceptibility and specific
heat measurements \cite{Hemberger} are summarized in
figure~\ref{fig:Phase_Diagram}(d).

\begin{figure}
\includegraphics[width=7cm]{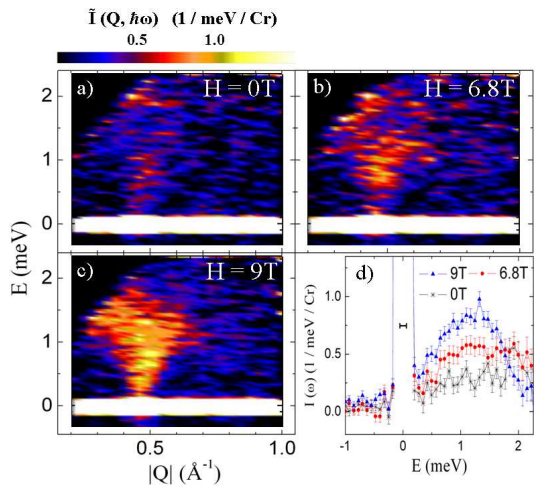}
\caption{\label{fig:DCS_H} \textbf{High field magnetic excitations:}
High resolution inelastic neutron scattering spectra at $T$=4 K and
(a) $H$=0 T, (b) $H$=6.8 T and (c) $H$=9 T, taken using 4.8 \AA\
incident neutrons. The background from the empty cryostat has been
subtracted off. (d) $\tilde{I}$($Q$,$\omega$) averaged over 0.3
\AA$^{-1}$$<Q<$0.6 \AA$^{-1}$. Horizontal bar shows instrument
resolution.}
\end{figure}

Figures~\ref{fig:DCS_H}(a)-(c) show $\tilde{I}(Q,\omega)$ spectra at
$T$=4 K in three external magnetic fields. These data employed 4.8
\AA\ incident neutrons which gives a reduced dynamic range relative
to Fig.~\ref{fig:DCS_T}. At $H$=0 T, weak low energy scattering from
residual helical ordered regions of the sample is again visible. As
the field is increased, this envelope of spin wave dispersion fills
in with inelastic intensity. By $H$=$H_c$ an excitation spectrum
characteristic of the purely helical ordered phase
(Fig.~\ref{fig:DCS_T}(b)) is recovered. Indeed, the total
fluctuating moment at $H$=9 T obtained by integrating over the
region $\hbar\omega\in[0.15,2]$meV and $Q\in[0.2,0.8]$\AA$^{-1}$ is
$\langle \delta m \rangle^{2}$=1.1(3)/Cr, comparable to the zero
field value measured in the purely helical phase at $T$=12 K. The
$Q$-averaged scattering (Fig.~\ref{fig:DCS_H}(d)) shows that the
magnetic field induces a shift of high energy spectral weight down
into a 1 meV maximum, in accordance with a conversion from collinear
to helical magnetic order. Since the magnetic inelastic scattering
cross section is proportional to the component of the fluctuating
moment perpendicular to $\hat{Q}$, strong inelastic scattering
together with a low value of $I(\vec{k}_1)$ at $H$=9 T is fully
consistent with the system being in a spin-flopped phase.

In summary, we have observed a rare field-induced commensurate to
incommensurate magnetic ordering transition \cite{Zheludev} in bond
frustrated ZnCr$_2$S$_4$, with a concomitant closing of a spin wave
gap. The phase boundary is fully consistent with that derived from
structural measurements, which indicates strong spin-phonon coupling
in this system. Recent theoretical work has shown that spiral
magnets, which break both time reversal and inversion symmetry,
permit a permanent electric dipole moment $\vec{P} \propto \vec{e}_3
\times \vec{k}$ where $\vec{e}_3$ is the spin rotation axis and
$\vec{k}$ the ordering wave vector \cite{Mostovoy}. While the
zero-field helical phase of ZnCr$_2$S$_4$ does not permit a finite
$\vec{P}$ since $\vec{e}_3
\parallel \vec{k}_1$, it would be interesting to search for an
electric polarization in the spin-flopped phase above $H_{SF}$ which
does permit one. Further progress towards refining the low
temperature magnetic structure in a field will require neutron
scattering experiments on single crystalline samples in both
vertical and horizontal field magnets.

We thank L. Li, K. Holman and Z. Tan for help with sample
characterization. We also acknowledge useful discussions with M. J.
Bhaseen, S. Sondhi, D. Huse and C. Broholm. The identification of any commercial product or trade name does not imply endorsement or recommendation by the National Institute of Standards and Technology.

\end{document}